\documentclass[pra,superscriptaddress,twocolumn,longbibliography]{revtex4-2}
\usepackage[T1]{fontenc}
\usepackage{amsmath}
\usepackage{amssymb}
\usepackage{mathtools}
\usepackage{bbm}
\usepackage{dsfont}
\usepackage[normalem]{ulem}
\usepackage{graphicx}
\usepackage{datetime}
\usepackage{color}
\usepackage{newtxtext,newtxmath}
\usepackage{microtype}
\usepackage{braket}
\usepackage{xr}
\usepackage{natbib}
\usepackage{hyperref}
\hypersetup{%
    bookmarksopen=false,
    bookmarksnumbered=true,
    pdfnewwindow=true,
    unicode=false,
    colorlinks=true,%
    citecolor=blue,
    linkcolor=black,
    urlcolor=blue,
    filecolor=blue
    }

\newcommand{\aDq}[1]{\big\langle\tilde{D}_{#1}\big\rangle}

\newcommand{\DE}{\Delta E} 
\newcommand{\eps}{\varepsilon}

\DeclareMathOperator{\var}{var}
\DeclareMathOperator{\Tr}{Tr}
\DeclareMathOperator{\spec}{Spec}
\newcommand{\vDq}[1]{\var\big(\tilde{D}_{#1}\big)}
\renewcommand{\vec}[1]{\boldsymbol{#1}}
\newcommand{\mms}{\rho_\textrm{MM}}
\newcommand{\HSD}{\mathcal{D}}
\begin{document}
\newcommand{\figdir}{.}
\newcommand{\liege}{Institut de Physique Nucl\'eaire, Atomique et de Spectroscopie, CESAM, Universit\'e de Li\`ege, B-4000 Li\`ege, Belgium}
\newcommand{\dlr}{German Aerospace Center, Institute of Quantum Technologies, D-89081 Ulm, Germany.}
\newcommand{\freiburg}{Physikalisches Institut, Albert-Ludwigs-Universit\"{a}t Freiburg, Hermann-Herder-Stra{\ss}e 3, D-79104, Freiburg, Germany}
\newcommand{\usal}{Departamento de F\'isica Fundamental, Universidad de Salamanca, E-37008 Salamanca, Spain}
\newcommand{\iffym}{Instituto Universitario de F\'isica Fundamental y Matem\'aticas (IUFFyM), Universidad de Salamanca, E-37008 Salamanca, Spain}
\newcommand{\eucor}{EUCOR Centre for Quantum Science and Quantum Computing, Albert-Ludwigs-Universit\"{a}t Freiburg, Hermann-Herder-Stra{\ss}e 3, D-79104, Freiburg, Germany}
\title{How to seed ergodic dynamics of interacting bosons under conditions of many-body quantum chaos}
\author{Lukas Pausch}
\altaffiliation[Present address: ]{\dlr}
\affiliation{\liege}
\author{Edoardo G. Carnio}
\affiliation{\freiburg}
\affiliation{\eucor}
\author{Andreas Buchleitner}
\email[]{a.buchleitner@physik.uni-freiburg.de}
\affiliation{\freiburg}
\affiliation{\eucor}
\author{Alberto Rodr\'iguez}
\email[]{argon@usal.es}
\affiliation{\usal}
\affiliation{\iffym}

\begin{abstract}
We demonstrate how the  initial state of ultracold atoms in an optical lattice controls the emergence of ergodic dynamics
as the underlying spectral structure is tuned into the quantum chaotic regime. Distinct initial states' 
chaos threshold values in terms of tunneling as compared to interaction strength are identified, as well as dynamical signatures of the chaos transition, on the level of experimentally accessible 
observables and time scales.
\end{abstract}
\maketitle
\section{Introduction}
\label{sec:intro}

Spectral and dynamical properties of closed quantum systems are intimately related to each other by the spectral decomposition of the unitary time 
evolution operator, in terms of the eigenenergies and -vectors of the underlying Hamiltonian \cite{hittmair72}. The actual imprint of the uniquely determined spectral structure on the
actually observed dynamics is, however, ultimately controlled 
by the decomposition of the specific initial state to be propagated in time, and of the measured observable, in the 
Hamiltonian's eigenbasis \cite{Geisel1986,Du1987,Brunner2023}. If the Hamiltonian features eigenstates with broadly variable localisation properties in 
Hilbert (or phase) space, distinct initial conditions can seed 
very distinct dynamical behaviour \cite{Giannoni89,abu1993,Carvalho2004,Brunner2017,Evrard2024}. In contrast, if all eigenstates delocalise alike, to homogeneously cover the energy surface
(apart from residual fluctuations -- 
which must secure their mutual orthogonality) \cite{Berry1977}, then
distinct initial states will rapidly relax into, at least on a coarse-grained level, qualitatively similar dynamical patterns, where 
interferences encoding the coherences between the initially populated eigenstates (and, thereby, the complete memory of the initial condition, which cannot be destroyed by unitary evolution,
however complex this may seem) average out, on sufficiently long time scales.

As a corollary, if a Hamiltonian's eigenstate structure can be continuously tuned 
from broadly distributed localisation properties to, in essence, uniform delocalisation, by an experimentally accessible control parameter, the dynamical evolution of suitably chosen 
initial states must unambiguously reflect these structural changes. While such tunable structural changes and the ensuing dynamics are qualitatively reasonably well 
understood within the theory of 
quantum chaos \cite{Giannoni89}, and reflect the structural metamorphoses of the underlying classical phase spaces, the latter remain largely uncharted when dealing with systems with 
more than two coupled degrees of freedom \cite{Tanner2000}. The associated phase spaces may exhibit intricate structural properties not available in the low-dimensional cases \cite{Gutzwiller1990}, 
and the reliable mapping of these structural properties \cite{Laskar1993,Milczewski1996,Schlagheck1999}, which shape the system's eigenstates, by analysis of the induced quantum dynamics turns ever more 
difficult with an increasing 
number of degrees of freedom \cite{Stoeber2024}.

Within the realm of cold atom physics, this scenario is precisely realised by interacting bosons loaded into optical lattices, described, in a most elementary way, by Bose Hubbard-like Hamiltonians \cite{Jaksch1998}: 
Each lattice site 
represents one degree of freedom, and the number $L$ of lattice sites, as well as the number $N$ of bosons per lattice site, can be controlled with high accuracy, over a broad range, such that 
the transition from single to many-body quantum dynamics can be continuously monitored, in phase spaces of controllable dimension \cite{Trimborn2009}. Furthermore, a single parameter allows to 
tune the Hamiltonian's eigenstate structure, at variable
$L>2$ and $N$, between the integrable and the 
ergodic (vulgo ``chaotic'') limit 
\cite{Buchleitner2003,Kolovsky2004,Ponomarev2006,Biroli2010b,Kollath2010,Beugeling2014,Beugeling2015,Beugeling2015c,Dubertrand2016,Beugeling2018,DelaCruz2020,Russomanno2020,Pausch2020,Pausch2021,Pausch2022,Kollath2007,Lauchli2008,Cramer2008a,Roux2009,Roux2010,Barmettler2012,Vidmar2013,Meinert2014a,Sorg2014,Andraschko2015,Despres2019,DelaCruz2020,WittmannW.2022}.
Due to the largely unexplored phase space 
structure, however, it is highly non-trivial
an issue which (experimentally preparable) initial states will trigger dynamics which faithfully reflect this metamorphosis, on experimentally explorable time scales.

Our present purpose therefore is
to provide the first quantitative, scalable (with $L$ and $N$) analysis of how the underlying eigenstate structure of Bose-Hubbard Hamiltonians is expressed in the dynamics of experimentally accessible initial states, probed by local (i.e., interrogating only a small subset of the system's degrees of freedom) observables.
More specifically, we
\emph{(i)} show that the dynamics of local observables ---within experimentally accessible time scales---  
can unambiguously pinpoint the chaotic phase, in remarkable agreement with the spectral 
characterization, \emph{(ii)} provide an analytical estimation of the onset of ergodic behaviour for arbitrary 
Fock initial states, and \emph{(iii)} illustrate how the initial 
state considered may affect the control parameter that drives the transition. We thus establish a scalable, quantitative benchmark of dynamical versus spectral properties of a 
paradigmatic quantum chaotic many-body system, opening a route towards robust control of large, strongly coupled, multi-component quantum systems with complex dynamics, 
an essential prerequisite for the design of quantum
computing platforms \cite{Berke2020,Basilewitsch2023,Borner2023}, and for the benchmarking of quantum simulators \cite{Choi2023,Mark2023}.

The manuscript is organized as follows: In Sec.~\ref{sec:model}, we introduce the physical model, and the family of initial
states considered 
in our analysis. The emergence of the spectrally chaotic phase, inferred
from an analysis of the localisation properties of the eigenvectors, 
along the initial states' trajectories in energy space
under 
variation of the Hamiltonian control parameter,
is presented in Sec.~\ref{sec:Dqresults}. The persistence of the chaotic domain as the thermodynamic limit is approached, and the definition of the control parameter driving the chaos
transition 
are discussed in Sec.~\ref{sec:control}, where we give 
an analytical estimate for the critical parameter value which defines the onset of chaos. In Sec.~\ref{sec:dynamics}, we finally benchmark the 
many-body dynamics seeded by the selected initial states against the underlying spectral and eigenvector structure, 
by studying the dynamics of
experimentally accessible local observables.
A summary of our findings is finally presented in Sec.~\ref{sec:conclusions}.

\section{Physical Model and Initial States}
\label{sec:model}
We study the one-dimensional, particle-number conserving Bose-Hubbard Hamiltonian (BHH) \cite{Lewenstein2007,Bloch2008,Cazalilla2011,Krutitsky2016} of $N$ bosons  
on $L$ sites,
\begin{align}
	H &= -J\sum_{j=1}^{L-1}\left(a^\dagger_j a_{j+1} + a^\dagger_{j+1} a_j\right) + \frac{U}{2} \sum_{j=1}^L a_j^\dagger a_j^\dagger a_j a_j, \label{eq:BHH}\\
	&\equiv -J h_\text{tun} + U h_\text{int}, \notag 
\end{align}
where $h_\text{tun}$ and $h_\text{int}$ denote, respectively, the dimensionless tunneling and interaction terms, $a_j^{(\dagger)}$ are creation and annihilation operators, associated with Wannier orbitals localized at each lattice site, and hard-wall boundary conditions (HWBC) are employed, as in 
typical experimental realizations 
with ultracold atoms in 
optical lattices. 

The Hamiltonian exhibits reflection symmetry about the center of the chain, i.e., $[H,\Pi]=0$, where $\Pi$ is the reflection operator. The Hilbert space 
of total  dimension $\HSD$ thus decomposes into the 
direct sum 
\begin{equation}
	\mathcal{H}=\mathcal{H}^+\oplus \mathcal{H}^-
\end{equation}
of the symmetric (even parity) 
and 
the antisymmetric (odd parity) subspace, $\mathcal{H}^+$ and $\mathcal{H}^{-}$ (of dimensions $\mathcal{D}^+$ and $\mathcal{D}^-$), respectively, which are invariant under the dynamics induced by $H$. 

In the limit of vanishing tunneling strength, $J=0$, as well as in the non-interacting limit, $U=0$, the BHH is integrable, 
and one 
finds as many independent and mutually commuting observables 
as the model's degrees of freedom, here given by the number $L$ of sites \cite{PauschThesis}. 
The eigenstates in those limits are thus Fock states uniquely identified by $L$ quantum numbers. For instance, the eigenstates in $\mathcal{H}^+$ for $J=0$ are
\begin{equation}
	\ket{\vec{n}^+} = \frac{1}{\sqrt{2(1+\delta_{\vec{n},\Pi\vec{n}})}}\left(\mathbbm{1} + \Pi \right) \ket{\vec{n}},
	\label{eq:FockStates}
\end{equation}
with $\ket{\vec{n}}\equiv\ket{n_1,\ldots,n_L}$, and $n_j$ the eigenvalues of the on-site number operators $\hat{n}_j = a^\dagger_j a_j$. 

When $J\neq 0$ and $U\neq 0$, the interplay of tunneling and interaction renders the BHH non-integrable and an ergodic
phase emerges as a function of $J$ and $U$, which can be 
identified from spectral and eigenvector features \cite{Kolovsky2004,Kollath2010,Dubertrand2016,Pausch2020,Pausch2021,Pausch2022}. 
In experiments with ultracold atoms, its 
existence 
is probed 
by analysing the dynamics of non-equilibrium initial configurations, which are typically Fock states in the on-site basis \cite{Cheneau2012,Meinert2014a,Meinert2014b,Kaufman2016,Rispoli2019,Lukin2018,Bohrdt2020,Takasu2020,Leonard2023,Trotzky2012,Bordia2016,Rubio-Abadal2019,Ronzheimer2013a,Choi2016a}. Most common initial states include \emph{(i)} the homogeneous density configuration \cite{Cheneau2012,Meinert2014a,Meinert2014b,Kaufman2016,Rispoli2019,Lukin2018,Bohrdt2020,Takasu2020,Leonard2023}, \emph{(ii)} staggered density distributions \cite{Trotzky2012,Bordia2016,Rubio-Abadal2019}, 
and \emph{(iii)} atomic clouds initially loaded into 
a strictly confined portion of the lattice, beyond which they can subsequently 
expand \cite{Ronzheimer2013a,Choi2016a}.

We here analyze the signature 
of the chaotic phase in the eigenstate decomposition and in the time evolution 
of exemplary 
Fock initial states of the above types, defined as:

\renewcommand\theenumi{\sl (\roman{enumi})}
\renewcommand\labelenumi{\theenumi}
\begin{enumerate}
	\item The \emph{homogeneous} density Fock state 
	\begin{equation}
		\ket{\psi_h} \equiv \ket{n,\ldots, n},
	\end{equation}
	for integer $n=N/L$. This configuration carries an enhanced weight of the system's ground state within the Mott insulator phase \cite{Lindinger2019}, and becomes the exact 
	ground state in the limit $J/U\to0$.
	
	\item The \emph{staggered} 
	configuration $\ket{\psi_s}$ 
	that at fixed density lies deepest in the spectrum bulk, (i.e., whose energy corresponds best to the arithmetic mean of the Hamiltonian's spectrum), and hence may exhibit maximal sensitivity to the emergence of spectral chaos. 
	For $n=1$, the corresponding Fock state has  
	$\lfloor (N-2)/3\rfloor$ sites occupied by 3 particles each, and 2 sites hosting 1 or 2 particles, to ensure the 
	desired $N$, 
	e.g., $\ket{\psi_s}=\ket{0203003020}$ for $N=L=10$ \cite{Note1}.
	Note that, at given density $n$, the staggered state's energy traces that of the maximally mixed (or ``infinite temperature") state $\rho_\text{MM}=\mathds{1}/\HSD$ (see Appendix \ref{ap:mms}),
	though the former state is pure, and thus exhibits coherences, what the latter does not.
	\item The \emph{localized} 
	initial state $\ket{\psi_\ell}$, where all $N$ particles are localised within 
	$\ell $ sites around the center of the lattice (we will here consider $\ell =3$ in all subsequent calculations), 
	each with $\left\lfloor N/\ell\right\rfloor$ or $\left\lceil N/\ell\right\rceil$ particles, for given
	$n$.
\end{enumerate}

\section{Parametric evolution of initial state energies}
\label{sec:Dqresults}

Let us now first analyse the spectral and eigenvector structure of the BHH, and its parametric evolution with variable $J$ and $U$, which we later want to 
certify through the observation of suitably chosen dynamical variables. For this latter purpose, we additionally need to consider which eigenstates are 
significantly populated by the choice of either one of the initial states (i-iii) above, as well as the parameter dependence of this decomposition.
Distinct decompositions will define distinct passages of the given initial states across the parameter space's chaotic domain, and thus lead to potentially
different experimental records -- triggering distinct interpretations.

We probe eigenvector ergodicity (in Fock space) using the finite-size generalized fractal dimension 
$\tilde{D}_1$,
\begin{align}
	\tilde{D}_1 = -\frac{1}{\log \HSD}\sum_{\alpha} |\psi_\alpha|^2 \log |\psi_\alpha|^2 \in [0,1],
\end{align}
where $\psi_\alpha$ is the amplitude of the state $\ket{\psi}$ on the basis state $\ket{\alpha}$ [e.g., the basis given in Eq.~\eqref{eq:FockStates}].
As we demonstrated in Refs.~\cite{Pausch2020,Pausch2021,Pausch2022,PauschThesis}, the chaotic region correlates 
with large $\tilde{D}_1$, converging towards 
unity in the limit $\HSD\to\infty$ 
[which corresponds to fully extended (ergodic) states in Hilbert space], and, most significantly, with 
strongly suppressed
eigenstate-to-eigenstate fluctuations of $\tilde{D}_1$. 

\begin{figure*}
	\includegraphics[width=.99\textwidth]{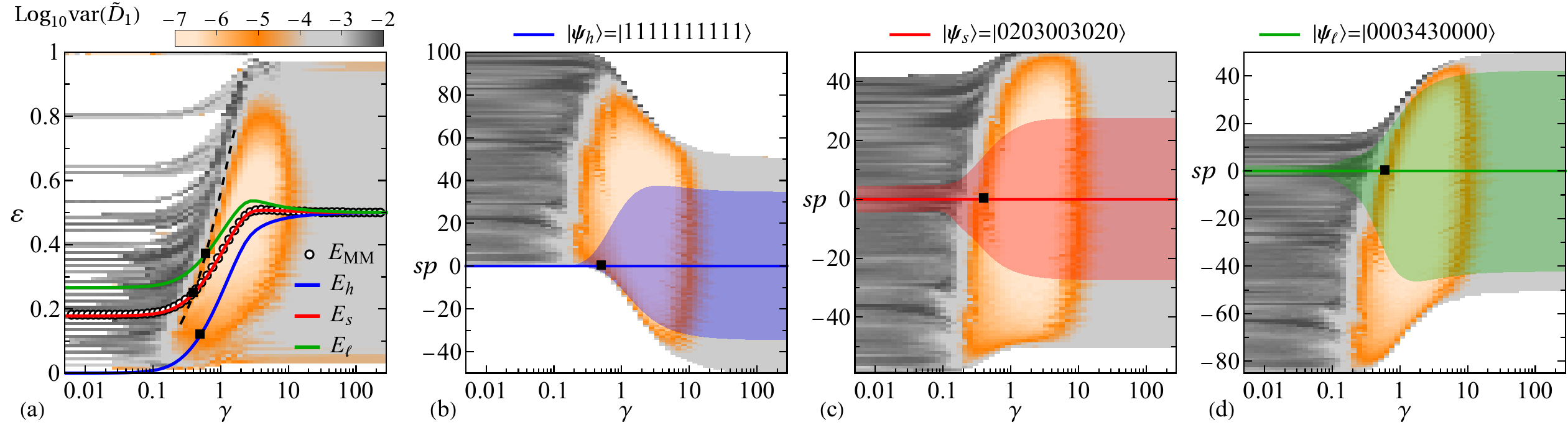}
	\caption{Perspectives of the chaotic phase for $L=N=10$ (HWBC, subspace $\mathcal{H}^+$, $\HSD^+=46\,252$) revealed by $\vDq{1}$ as a function of $\gamma=J/U$ and (a) the scaled energy $\varepsilon=(E-E_\text{min})/(E_\text{max}-E_\text{min})$ or (b)-(d) the spectrum percentage ($sp$) measured from the reference energy of the indicated Fock states. For each $J/U$ value, the eigenstates are sorted  into 100 sets that (a) correspond to intervals of constant width in $\varepsilon$ or (b)-(d) contain the same number of states (i.e., same spectrum percentage), and $\vDq{1}$ is evaluated for each set. Solid lines in (a) show the energy trajectories of the selected Fock states [Eqs.~\eqref{eq:E_h}-\eqref{eq:E_l}], while symbols correspond to the maximally mixed state [Eq.~\eqref{eq:Emms}]. Shaded regions in (b)-(d) mark the energy width $\pm\sigma$ [Eq.~\eqref{eq:sigmaE}] of each Fock state in terms of $sp$. Black squares indicate the threshold values $\gamma^c$ from Eq.~\eqref{eq:gammacH} ($\ket{\psi_h}$) and  Eq.~\eqref{eq:gammac} ($\ket{\psi_s}$, $\ket{\psi_\ell}$), which gives the dashed trajectory in (a).}
	\label{fig:DensityPlotsVarDq}
\end{figure*}

Figure \ref{fig:DensityPlotsVarDq}(a) shows the chaotic phase for $N=L=10$, exposed by the suppressed variance of $\tilde{D}_1$ over close-in-energy eigenstates, in terms of the 
relative tunneling strength 
\begin{equation}
  \gamma\equiv J/U \, ,
\end{equation}
and of the 
scaled 
energy \cite{Pausch2020}
\begin{equation}
 \eps = \frac{E - E_\text{min}}{E_\text{max} - E_\text{min}}\, ,
\end{equation}
where $E_{\text{min}(\text{max})}$ denotes the minimum (maximum) eigenenergy for each $J/U$. 
We also show the initial states' (i-iii) trajectories in energy space under variation of $\gamma$, as determined by 
\begin{alignat}{2}
	\label{eq:E_h}
	E_h &\equiv \bra{\psi_h} H \ket{\psi_h} && = \frac{UN}{2}(n-1), \\
	\label{eq:E_s}
	E_s &\equiv \bra{\psi_s} H \ket{\psi_s} && =  U (N-2),\\
	\label{eq:E_l}
	E_\ell &\equiv \bra{\psi_\ell} H \ket{\psi_\ell} && =  \frac{U}{2\ell}\left(N-r_{N/\ell}\right)\left(N+r_{N/\ell}-\ell\right), 
\end{alignat}
where $r_{N/\ell}\equiv (N\mod \ell) \in[0,\ell-1]$, and, by choice, 
$E_s$ approximates the energy
\begin{alignat}{2}
  E_\text{MM} &\equiv \Tr\left(H\rho_\text{MM}\right) &&= U \frac{N(N-1)}{L+1} \notag \\
  & &&= U\left[nN -n(n+1) +O(N^{-1})\right]
  \label{eq:Emms}
\end{alignat}  
of the maximally mixed state $\rho_\text{MM}$, 
with $n=1$ (see Appendix \ref{ap:mms} for the 
derivation of $E_\text{MM}$).

As can be observed in Fig.\ref{fig:DensityPlotsVarDq}(a), the Fock initial states' trajectories 
do not remain at a constant level of 
the scaled energy. 
Instead, 
$\varepsilon$ varies strongly with $\gamma$, along each 
trajectory,
and the chaotic phase is entered at different points by the different initial states' trajectories, before they all 
converge into the same limiting 
value $\eps=0.5$, for  
$\gamma\to\infty$ \cite{Note2}
Hence, chaos 
emerges differently along the energy trajectory of each Fock initial state, 
and, consequently, in the associated dynamical behaviour (as we will confirm further down).

Panels (b)-(d) in Fig.~\ref{fig:DensityPlotsVarDq} present the chaotic phase, in terms of $\vDq{1}$, as a function of $\gamma$, and of the 
percentage of the spectrum that lies above and below the given initial 
state's energy, for $N=L=10$ and $\ell=3$.
Once the region of uniformly delocalised eigenstates, 
characterised by a strongly suppressed
variance $\vDq{1}$ of the generalised fractal dimension emerges, it practically extends over
the entire spectrum.

Also shown is 
the evolution of the energy width $\pm \sigma$ of each 
Fock initial state, in terms of the spectral percentage, given by
\begin{align}
 \sigma^2_{\ket{\vec{n}}} &\equiv \braket{H^2}_{\ket{\vec{n}}}-\braket{H}_{\ket{\vec{n}}}^2 \notag\\
 &=J^2\left(2N-n_1-n_L+2\sum_{j=1}^{L-1} n_jn_{j+1}\right) \, ,
 \label{eq:sigmaE}
\end{align}
for $\ket{\vec{n}}=\ket{\psi_{h,s,\ell}}$.
This energy scale quantifies the width of the 
local density of states (LDOS) of $H$ with respect to $\ket{\psi_{h,s,\ell}}$,
and hence 
of the spectral region where eigenstates of $H$ noticeably contribute to 
$\ket{\psi_{h,s,\ell}}$.

The plots thus illustrate 
how the chaotic phase unfolds from the perspective of the different initial states 

$\ket{\psi_h}$ (b), $\ket{\psi_s}$ (c), and $\ket{\psi_\ell}$ (d):
As $\gamma$ is increased from the strongly interacting limit, the entrance of each Fock state into the chaotic 
domain correlates with a pronounced increase of $\sigma$ in terms of the spectral percentage, consistent with the emerging
ergodic structure of chaotic eigenstates in Fock space. The number of eigenstates that contribute dominantly to each 
Fock initial state reaches its maximum 
around the centre of the chaotic phase, and barely changes as the non-interacting limit is approached. 
In the $\gamma$ range where each 
initial state picks up 
significant
contributions from 
ergodic eigenstates, 
the dynamical manifestations of chaos 
must become observable.

\section{Persistence of ergodicity in the thermodynamic limit}
\label{sec:control}

\begin{figure}
	\includegraphics[width=\columnwidth]{\figdir/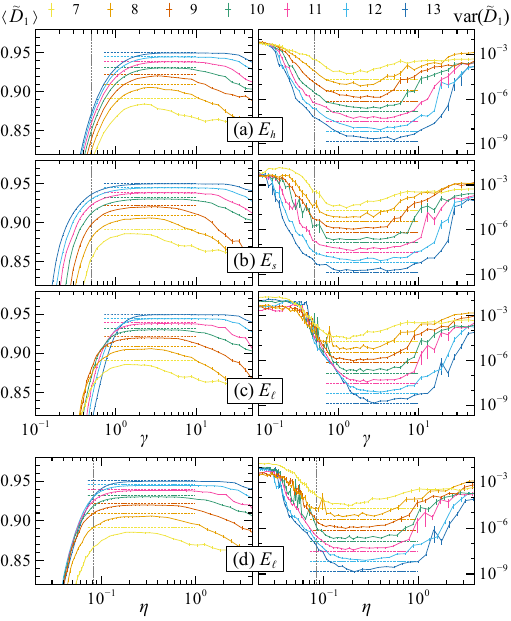}
	\caption{Dependence on $\gamma=J/U$ (top three rows) and on $\eta=J/UN$ (bottom row) of $\aDq{1}$ (left column) and of $\vDq{1}$ (right column) 
	calculated from $\gtrsim 100$ eigenstates (subspace $\mathcal{H}^+$) closest in energy to (a)
	the homogeneous density state $\ket{\psi_h}$, (b) the staggered density state $\ket{\psi_s}$, and (c,d) the localized density state $\ket{\psi_\ell}$
	for varying values of $N=L\in[7,13]$ with $\mathcal{D}^+\in[868,2\,600\,612]$.
	Horizontal dashed lines  
	indicate GOE predictions 
	for the corresponding Hilbert space size. Vertical dashed lines mark the threshold values $\gamma^c$ for $N\to\infty$ from Eqs.~\eqref{eq:gammacH} and \eqref{eq:gammac}, respectively for (a) and (b), and $\eta^c$ from Eq.~\eqref{eq:etac} for (d). 
	}
	\label{fig:JU_JUN_Comparison}
\end{figure}

In the previous section, we analysed the passage of the initial state trajectories across energy space, under variation of $\gamma$, and at fixed particle number $N$ and density $n$.
Since quantum chaos and also random matrix theory explore quantum signatures \cite{Haake2018} of classical mechanics and thermodynamics, it is important to clarify how the essential
features identified in such analysis evolve with the system size, and in particular in the thermodynamic limit (TL),
i.e., for $N\to\infty$, fixed 
$n$ \cite{Abufootnote}.

To monitor this evolution with the initial states (i-iii) as reference, 
Figure~\ref{fig:JU_JUN_Comparison} shows the mean and the variance of $\tilde{D}_1$ over the $\gtrsim 100$ eigenstates closest in energy to (a) $E_h$, (b) $E_s$, and (c) $E_\ell$ [and therefore 
certainly within their respective energy widths $\sigma$ as indicated in Figs.~\ref{fig:DensityPlotsVarDq}(b-d)], 
as functions of $\gamma$, 
and for $N\in[7,13]$ and $n=1$.
That number of eigenstates is small enough to probe the properties of the spectrum only locally and large enough to ensure good statistics for $\aDq{1}$ and $\vDq{1}$.
In all three cases, the ergodic phase is revealed by plateaux of maximal $\aDq{1}$ (converging to unity, in the TL), correlating with minimal $\vDq{1}$ (converging to 0, in the TL), which are both
well described by the corresponding values obtained from the Gaussian Orthogonal Ensemble (GOE) of random matrices \cite{Pausch2020,Pausch2021,Pausch2022}.
As the system size is increased, the ergodic phase in the spectral vicinity of 
the homogeneous $\ket{\psi_h}$ [panel (a)] and of the staggered density state $\ket{\psi_s}$ [panel (b)] is getting ever more clearly exposed, within a well-defined range of 
$\gamma$ values. 
In contrast, the emergence of chaos 
in the spectral vicinity of the localized density state $\ket{\psi_\ell}$ [panel (c)] 
systematically shifts to 
larger $\gamma$ with increasing system size [note the 
consistent displacement of the  $\aDq{1}$ curves and of the minimum in $\vDq{1}$]. Only when the tunneling strength is scaled as
\begin{equation}
 \eta\equiv J/UN
 \label{eq:eta}
\end{equation}
[panel (d)] is this shift removed and the appearance of the chaotic region eventually becomes system-size independent. 
While this may appear intuitive, since increasing 
$N$ at fixed $n$ {\em and} $\ell$ 
increases the effective interaction strength for $\ket{\psi_\ell}$, such that also $\gamma$ needs to be increased to compensate this effect, we will now make this argument more 
quantitative, by closer inspection of 
the associated 
scaling properties of the spectrum. This will allow us to extract the critical values of $\gamma$ and $\eta$, respectively, at which the chaos transition occurs for the different initial states.

In general, for a chaotic spectral and eigenstate structure to persist, 
the tunneling and the interaction terms in $H$ ---which represent incompatible integrals of motion---
need to balance each other: If either of them became progressively dominant as the boson number 
is increased, the system would be drawn into the 
corresponding 
integrable 
phase. It is therefore instructive to consider the scaling behaviour of the tunneling and of the interaction 
contribution to $H$: 
The spectral width, 
\begin{equation}
 \Delta E\equiv E_\text{max}-E_\text{min},
\end{equation}
of the tunneling term reads $\DE_\text{tun}=4JN$ and, hence, scales linearly with $N$. On the other hand, at integer density $n$, the spectral width $\DE_\text{int}=UN(N-n)/2$ 
of the interaction term 
scales quadratically with $N$.
The interaction term will therefore induce a shift $\propto N^2$ 
of some eigenenergies 
when the TL is approached at constant particle density.

\begin{figure}
	\centering
	\includegraphics[width=\linewidth]{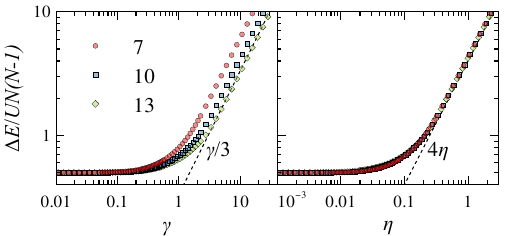}
	\caption{Spectral width $\DE=E_\text{max}-E_\text{min}$ of $H$ scaled by $UN(N-1)$ versus $\gamma=J/U$ (left) and $\eta = J/UN$ (right), for $N=L\in[7,13]$ as indicated. 
	Dashed lines denote the exact linear dependence of the scaled spectral width in the limits 
	$\gamma\to\infty$ (for $N=13$) and $\eta\to\infty$ (for $N\to\infty$).
	}
	\label{fig:SpectralWidth}
\end{figure}

This is demonstrated in 
Fig.~\ref{fig:SpectralWidth}, for $n=1$, via the dependence of the scaled spectral width $\DE/UN(N-1)$ on $\gamma$, $\eta$, and $N$.
In the strong interaction regime ($\gamma\, ,\, \eta \to 0$), the scaled spectral width converges to $1/2$, while it approaches the asymptotes $4\gamma/(N-1)$ and $4\eta$ in the tunneling-dominated regime, i.e. for 
$\gamma\gg 1$ and $\eta \gg 1$ 
(provided $N\gg 1$, in the latter case), respectively.
Notably, the scaled spectral width becomes system-size independent as a function of $\eta$, and the crossover between both regimes can be located in terms of the rescaled 
tunneling strength at $\eta_*\simeq 1/8$.
 Since, for $\eta<\eta_*$, there is a subset of eigenenergies which scale, at fixed $n$ and $\gamma$, 
quadratically with $N$, the interaction term will progressively dominate their character as $N$ increases, and they must exit the chaotic phase, in the TL.

The energy expectation values [Eqs.~\eqref{eq:E_h}-\eqref{eq:E_l}] of the Fock initial states (i-iii) are, by construction, entirely determined by the interaction term in (\ref{eq:BHH}), and these scale,
at fixed $n$, as $E_{h,s}\sim N$ and $E_\ell\sim N^2$. This different scaling reflects the initial states' structure, which for $\ket{\psi_{h,s}}$ explores the entire lattice (which grows with $N$, since $n$ is kept 
constant in the TL), while for $\ket{\psi_\ell}$ concentrates
ever more particles on $\ell$ sites, thus activating the inter-particle interaction term. Consequently, at fixed $\gamma$ and $n$, $\ket{\psi_\ell}$ will remain interaction-dominated at sufficiently large 
$N$, and hence exit the chaotic domain, as visible in Fig.~\ref{fig:JU_JUN_Comparison}(c). Only when this concentration effect is compensated by rescaling $U$ with $N$ 
does the extent of the ergodic phase for $\ket{\psi_\ell}$ remain well-defined in the TL, in terms of $\eta$ [see Fig.~\ref{fig:JU_JUN_Comparison}(d)].
This is in line with studies of the semiclassical limit of the BHH 
\cite{Castro2024,Dubertrand2016,Hiller2006,Hiller2009}, 
where the relevant system parameters determining the emergence of chaos are $\eta$ and $E/UN^2$, and the limit $n\to\infty$ is performed 
at constant $L$---reminiscent of our present scenario for $\ket{\psi_\ell}$, locally on the $\ell$ initially occupied sites.

%
Given the above scaling arguments, for chaos to prevail in the TL for given Hamiltonian parameters $J$ and $U$,
this limit needs to be approached at constant energy density $\omega\equiv \langle H \rangle/N$.
At well-defined $\omega$, however, also an appropriate choice of $\gamma$ is required to balance the contributions of the tunneling and interaction terms in $H$.

\begin{figure}
	\centering
	\includegraphics[width=\linewidth]{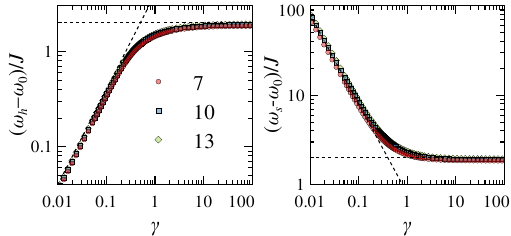}
	\caption{
		Energy density excess $(\omega-\omega_0)/J$
		as a function of $\gamma= J/U$ for $\ket{\psi_h}$ (left) and $\ket{\psi_s}$ (right). Dashed lines show the asymptotic tendencies given in Eq.~\eqref{eq:AsymExcessEdensity} (with $N=10$ for $\ket{\psi_s}$).}
	\label{fig:ExcessEdensity}
\end{figure}
For a given $\omega$, one may obtain a rough estimate of the $\gamma$ value at which the chaotic domain appears by analysing the behaviour of $(\omega -\omega_0)/J$ as 
a function of $\gamma$, where $\omega_0$ corresponds to the Hamiltonian's ground state's energy density. For Fock states, 
\begin{equation}
 \frac{\omega}{J}=\frac{1}{\gamma} \frac{\langle h_\text{int}\rangle}{N},
\end{equation}
while, at integer density, 
\begin{alignat}{2}
 \frac{\omega_0}{J} &= \frac{n-1}{2\gamma}-2(n+1)\gamma + O(\gamma^3), &\quad &\gamma\to 0,\\
 \frac{\omega_0}{J} &= -2 +O(1/\gamma), &\quad &\gamma\to \infty,
\end{alignat}
as follows from standard perturbation theory. Therefore, asymptotically, one has 
\begin{subequations}
\begin{align}
 \frac{\omega-\omega_0}{J} &\underset{\gamma\to 0}{=} \frac{1}{\gamma}\left(\frac{\langle h_\text{int}\rangle}{N} -\frac{n-1}{2}\right) +2(n+1)\gamma,\\
 \frac{\omega-\omega_0}{J} &\underset{\gamma\to \infty}{=} 2,
\end{align}
\label{eq:AsymExcessEdensity}
\end{subequations}
where two terms are kept in the limit $\gamma\to0$ to account for the case where $\langle h_\text{int}\rangle/N=(n-1)/2$, as it precisely happens for $\ket{\psi_h}$.
The above two limiting behaviours govern the evolution of the energy density excess $(\omega -\omega_0)/J$ in terms of $\gamma$, as demonstrated in Fig.~\ref{fig:ExcessEdensity} for $\ket{\psi_h}$ 
and $\ket{\psi_s}$. The crossing point of both asymptotic tendencies 
may be used as an estimate of the threshold value $\gamma^c$ at which the tunneling term starts taming the interaction contribution, and hence of the region where one 
expects chaos to emerge. According to Eqs.~\eqref{eq:AsymExcessEdensity}, for the homogeneous state, which happens to be the Fock state with the lowest 
possible energy density, $\omega/J=(n-1)/2\gamma$,  the crossing point reads 
\begin{equation}
 \gamma^c_h=\frac{1}{n+1},
 \label{eq:gammacH}
\end{equation}
and, for a Fock state with higher energy density, 
\begin{equation}
 \gamma^c = \frac{1}{2}\left(\frac{\langle h_\text{int}\rangle}{N}-\frac{n-1}{2}\right).
 \label{eq:gammac}
\end{equation}
For $\ket{\psi_h}$ and $\ket{\psi_s}$ at unit density, one has $\gamma^c_h=1/2$, and $\gamma^c_s=(1-2/N)/2\approx1/2$ for large $N$, 
which, as indicated in Figs.~\ref{fig:JU_JUN_Comparison}(a)-(b) and Figs.~\ref{fig:DensityPlotsVarDq}(b)-(c), describe qualitatively well the threshold of the region with ergodic 
eigenstates, and the entrance of these 
Fock initial states into the chaotic domain. The estimate \eqref{eq:gammac} may also be applied to the case where $\langle h_\text{int}\rangle\sim N^2$. In this 
case, one obtains, at constant density and for large $N$,
\begin{equation}
 \eta^c\simeq\frac{1}{2}\frac{\langle h_\text{int}\rangle}{N^2},
 \label{eq:etac}
\end{equation}
which for $\ket{\psi_\ell}$ 
captures fairly well the appearance of the chaotic regime as the TL is approached, as observed in
Fig.~\ref{fig:JU_JUN_Comparison}(d).  Furthermore, the relation \eqref{eq:gammac} between the dimensionless energy density $\langle h_\text{int}\rangle /N$
and the threshold value $\gamma^c$, when visualized in terms of the scaled energy $\varepsilon$ for fixed $N$, describes correctly the tilt of the chaotic phase towards 
higher relative tunneling strengths for larger scaled energies, as 
indicated in Fig.~\ref{fig:DensityPlotsVarDq}(a).

\section{Dynamical behaviour of experimentally accessible local observables}
\label{sec:dynamics}
Our above
analysis of the spectral and eigenvector properties of $H$ led
to an 
estimate 
for the chaos threshold of 
different Fock states $\ket{\psi_{h,s,\ell}}$, and 
the scaling of the states' energy densities when approaching the TL allowed us 
to conclude whether such threshold correlates with well-defined threshold values $\gamma^c$ or $\eta^c$,  
for varying system sizes.

Let us now investigate how 
this spectral 
picture translates into dynamical features.
For this purpose, we calculate the time evolution of the initial states $\ket{\psi_h}$, $\ket{\psi_s}$, and $\ket{\psi_\ell}$, for increasing system size, 
at constant density $n=1$, up to a maximum of 200 tunneling times, $\tau \equiv tJ/\hbar <200$, chosen to lie within time scales experimentally accessible with ultracold bosons, 
as those probed in Refs.~\cite{Rispoli2019,Leonard2023}. The dynamical 
evolution is 
numerically generated by means of an expansion of the unitary time evolution operator in terms of Chebyshev polynomials, which permits to efficiently simulate the dynamics in large 
Hilbert 
spaces (see Appendix \ref{ap:chebyshev} for technical details). For simplicity, we will restrict ourselves to observables that can be easily reconstructed from uncorrelated measurements 
of single-site densities, and which are thus within 
reach for state-of-the-art experiments.

\begin{figure}
	\centering
	\includegraphics[width=\columnwidth]{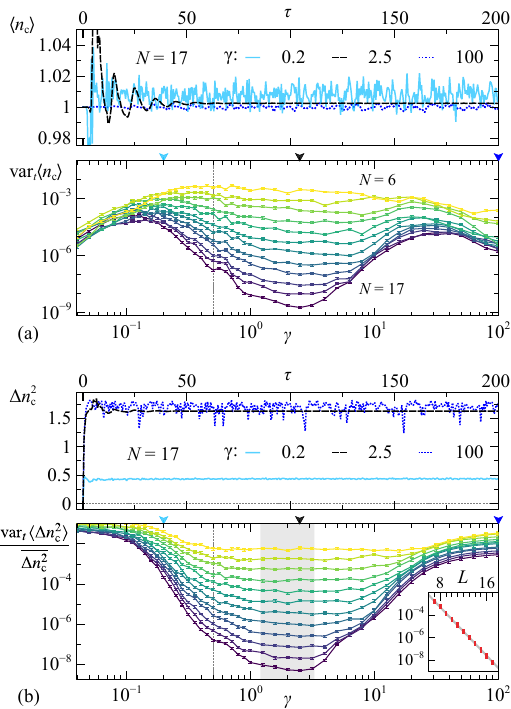}
	\caption{Dynamical features of (a) on-site density expectation values $\left\langle n_\text{c}(\tau)\right\rangle$ and (b) the fluctuations $\Delta n_\text{c}^2(\tau)$ on a central site (as explained in the main text),
	for the initially homogeneous configuration $\ket{\psi_h}$. 
	The upper panel in each subfigure shows the time evolution for $N=17$ and three values of $\gamma$ as indicated 
	(marked in the lower panels by correspondingly coloured arrowheads). 
	Each lower panel displays 
	the (relative) time variance of the corresponding signal 
	as a function of $\gamma$ for $N=L\in[6,17]$ (from top to bottom). 
	The vertical dashed lines indicate $\gamma_h^c=1/2$ [Eq.~\eqref{eq:gammacH}]. The inset in (b) shows the relative time variance averaged in the shaded area versus $L$, where the solid line is the best fit  $12.15 e^{-1.25L}$.}
	\label{fig:DynamicsH}
\end{figure}
\subsection{Homogeneous Fock state}
We start out with 
the homogeneous initial configuration, $\ket{\psi_h}$. The top panels of Figs.~\ref{fig:DynamicsH}(a,b) 
show, respectively, the time evolution of the on-site density expectation value $\left\langle n_\text{c}(\tau)\right\rangle$, and of the corresponding fluctuations 
$\Delta n_\text{c}^2(\tau)\equiv \langle \left[n_\text{c}(\tau)-\langle n_\text{c}(\tau)\rangle \right]^2\rangle$,  
measured at a central site with index $\lfloor L/2\rfloor$, i.e., taken to be the closest site (from the left) to the lattice centre. This choice permits to treat even and odd system sizes on an equal footing.
Given the initially homogeneous density, no net mass transport in time across the system can be expected in the TL, and, hence, on-site densities for finite $L$ should merely exhibit small 
variations around the value $n=1$, induced by the presence of hard-wall boundaries, as observed in Fig.~\ref{fig:DynamicsH}(a) for $N=17$ and three representative values of $\gamma$. 
For $\gamma=2.5$, which is deep in the chaotic regime, according to our above spectral
analysis, initial oscillations in $\left\langle n_\text{c}(\tau)\right\rangle$ fade out on the time scale of $\tau\approx 50$ tunneling times, and the central site 
then equilibrates, settling at a sharp value of $\langle n_\text{c}\rangle$ \cite{Note3}.
In contrast, temporal fluctuations 
persist throughout the 
entire time window for the other two $\gamma$ values, representing the interaction-dominated and tunneling-dominated regimes.
An analogous behaviour is found 
for the on-site number variance  
$\Delta n_\text{c}^2(\tau)$ in Fig.~\ref{fig:DynamicsH}(b), 
which again settles at a sharp value, for $\gamma = 2.5$ and $\tau \geq 50$, while outside the chaotic domain also this two-particle observable (sensitive to two-particle interference 
contributions \cite{Brunner2017}) exhibits persistent fluctuations.

We quantify the time fluctuations
of the expectation value $\langle O(\tau)\rangle$ of a given observable $O$ by its 
temporal variance in the interval $\tau\in[\tau_\text{i},\tau_\text{f}]$,
\begin{align}
	\var_t \langle O \rangle = \frac{1}{\tau_\text{f}-\tau_\text{i}}\int_{\tau_\text{i}}^{\tau_\text{f}} \mathrm{d}\tau \left(\langle O(\tau)\rangle - \overline{\langle O(\tau)\rangle}\right)^2,
\end{align}
where the overline  
denotes the time average 
within the same time interval. The lower panels in Figs.~\ref{fig:DynamicsH}(a,b)  
show, respectively, the temporal variance for $\tau\in[100,200]$ of the signals $\left\langle n_\text{c}(\tau)\right\rangle$ and 
$\Delta n_\text{c}^2(\tau)$ (the relative variance in this latter case, i.e., scaled by $\overline{\Delta n_\text{c}(\tau)^2}$), as functions of $\gamma$, for varying system size. 
For the local one- and two-particle observables considered, a distinctive regime of markedly suppressed temporal fluctuations emerges in the range $0.2\lesssim \gamma \lesssim 20$ for 
all system sizes, which correlates unambiguously with the chaotic phase identified in the preceding spectral and eigenvector 
analysis [cf.~Figs.~\ref{fig:DensityPlotsVarDq}(b) and \ref{fig:JU_JUN_Comparison}(a)]. The analytically obtained value $\gamma_h^c=1/2$ also captures correctly the threshold 
of the region with minimal temporal fluctuations, which are ever reducing for increasing system size. 
As demonstrated in the inset to Fig.~\ref{fig:DynamicsH}(b), these fluctuations decrease exponentially with $L$ (i.e., 
with the number of degrees of freedom of the system \cite{PauschThesis}), in accordance with the 
emergence of an ergodic dynamical regime \cite{Srednicki1996,Srednicki1999}. 

These observations then confirm $\gamma$ as the parameter that controls the dynamical appearance of the chaotic phase for $\ket{\psi_h}$, as deduced from the energy scaling in the previous section. 

\subsection{Staggered density configuration}
Let us now study the dynamics of initial Fock states $\ket{\psi_s}$ corresponding to staggered density configurations. As discussed in Sec.~\ref{sec:model}, these states, 
given in Table \ref{tab:StagStates} for different $L$, lie deep in the spectrum bulk, their energies being closest to the arithmetic average of the Hamiltonian's spectrum [cf.~Eqs.~\eqref{eq:E_s} and \eqref{eq:Emms}].
\begin{table}
 \begin{tabular}{ll}
  \hline\hline  
  $L$ & $\ket{\psi_s}$ \\\hline 
  11 & $\ket{01030303010}$ \\ 
  13 & $\ket{0200303030020}$ \\
  14 & $\ket{01030300303010}$ \\
  16 & $\ket{0200303003030020}$ \\
  17 & $\ket{01003030303030010}$\\
  \hline\hline 
 \end{tabular}
\caption{Initial Fock states at unit density describing staggered density configurations whose energy corresponds best to the arithmetic average of the Hamiltonian's spectrum 
[cf.~Eqs.~\eqref{eq:E_s} and \eqref{eq:Emms}].}
\label{tab:StagStates}
\end{table}
\begin{figure*} 
	\centering
	\includegraphics[width=.98\textwidth]{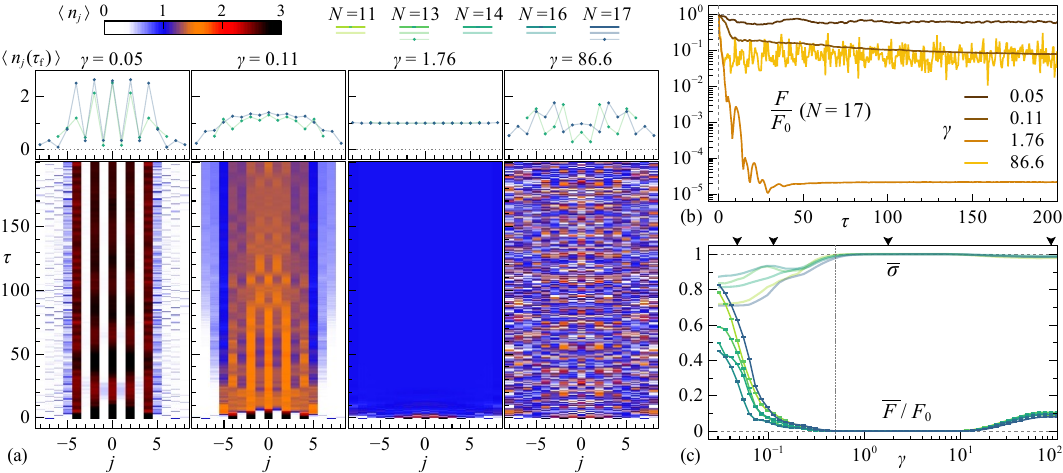}
	\caption{Dynamical observables for the initially staggered density configurations given in Table \ref{tab:StagStates}. Density plots in (a) show the evolution of on-site density expectation values $\langle n_j(\tau)\rangle$ with tunneling time $\tau$ for $N=17$ and the four values of $\gamma$ indicated [also marked by arrowheads in panel (c)], and are topped by the spatial density profiles at the final evolution time $\tau_\textrm{f}=200$. The time evolution of the density mean square deviation $F(\tau)$ [Eq.~\eqref{eq:F}, scaled by $F_0\equiv F(\tau=0)$] corresponding to the density plots is displayed in panel (b). The time-averaged value of the density profile width [Eq.~\eqref{eq:sigmaW}], $\overline{\sigma}$, and of the density mean square deviation, $\overline{F}$, in the interval $\tau\in[100,200]$ are shown as functions of $\gamma$ for varying system sizes (color coded) in panel (c). The vertical dotted line in (c) marks the threshold value $\gamma^c_s=1/2$ from Eq.~\eqref{eq:gammac} for $N\to\infty$.}
	\label{fig:DynamicsStag}
\end{figure*}

In this case, given the non-homogeneous nature of the initial state, useful information can be gathered by monitoring the evolution of the spatial density 
profile (defined by the on-site density expectation values $\langle n_j(\tau)\rangle$), with an example displayed 
for $N=17$ and four representative $\gamma$ values in Fig.~\ref{fig:DynamicsStag}(a). 
For $\gamma=0.05$, the large value of the interaction energy induces self-trapping, and the bosons remain localized on initially multiply occupied sites, flanked by oscillations of two single particles at the lattice ends. Traces of the initial condition thus remain well visible at the final time $\tau_\text{f}=200$ [see top panels in Fig.~\ref{fig:DynamicsStag}(a)]. 
Increasing the relative tunneling strength, for $\gamma=0.11$, we observe 
a progressive melting of the staggered pattern and the formation of a density cloud that falls off towards the edges. Deep in the chaotic phase as demarcated in our spectral
analysis, 
for $\gamma=1.76$, the density profile 
homogenizes very quickly, and after $\approx 10$ tunneling times the density distribution remains equilibrated in time, 
without any manifestation of the many-particle dynamics' coherent nature.
When approaching the non-interacting limit, for $\gamma=86.6$, the entire evolution is strongly imprinted 
by interference effects, and the spatial density distribution, although arguably 
more homogeneous than initially, exhibits persistent temporal fluctuations.

The emergence of many-body quantum chaos entails a fast temporal growth of the entanglement among multiple degrees of freedom (DOF -- here associated with the individual lattice sites) 
of the system. The reduced density matrices $\rho_\text{red}$ of small subsystems (i.e., involving reduced DOF subsets) will thus quickly develop a strong mixedness, and, 
correspondingly, the expectation values of local observables (living on a subset of lattice sites), which would be entirely determined by a state of the form $\rho_\text{red}$, 
dephase rapidly in time and converge to a steady state value, as reflected by the single-particle expectation values $\langle n_j(\tau)\rangle$ for $\gamma=1.76$ (and the total 
absence of interference-induced patterns) in Fig.~\ref{fig:DynamicsStag}(a).

To identify dynamically the chaotic phase from the behaviour of the density profile, we 
characterize the spatial width of the atomic cloud by 
\begin{equation}
  \sigma(\tau) = \frac{1}{\sigma_\text{u}}\left[\sum_{j=1}^L \frac{\langle n_j(\tau)\rangle}{N} \left(j-\frac{L+1}{2}\right)^2\right]^{1/2},
  \label{eq:sigmaW}
\end{equation}
which gives the standard deviation of the discrete distribution $\langle n_j(\tau)\rangle/N$, $j=1,\ldots,L$, normalized with respect to the value for a discrete uniform distribution, 
$\sigma_\text{u}=\sqrt{(L^2-1)/12}$. We further quantify the cloud's homogeneity by the mean square deviation of the on-site density values from $n$,  
\begin{align}
  F(\tau) = \frac{1}{L}\sum_{j=1}^{L} \left[\langle n_j(\tau)\rangle - n\right]^2,
  \label{eq:F}
\end{align}
which vanishes for a uniform density distribution. 

The evolution of $F(\tau)$ shown in Fig.~\ref{fig:DynamicsStag}(b) for $N=17$ encodes the time development of the density profiles of Fig.~\ref{fig:DynamicsStag}(a): While 
deviations from homogeneity are notable for low and high relative tunneling strengths, $F(\tau)$ drops quickly by several orders of magnitude for $\gamma=1.76$. 

The time-averaged 
values, over the interval $\tau\in[100,200]$, of the figures of merit \eqref{eq:sigmaW} and \eqref{eq:F}, as functions of $\gamma$, and for varying system size, are displayed
in Fig.~\ref{fig:DynamicsStag}(c). There, we 
identify a parametric range $0.5\lesssim \gamma\lesssim 10$ where $\overline{F}$ goes to zero as the width $\overline{\sigma}$ approaches unity, which is common for all system sizes, 
correlates unambiguously with the extension of the chaotic phase as inferred from the 
spectral
analysis [cf.~Figs.\ref{fig:DensityPlotsVarDq}(c) and \ref{fig:JU_JUN_Comparison}(b)], and its appearance is well captured by the analytical 
estimation from Eq.~\eqref{eq:gammac}, $\gamma^c_s= 1/2$ $(N\to \infty)$. 
It is worth mentioning that the $\gamma$-dependence of the temporal fluctuations of $\sigma(\tau)$ and $F(\tau)$ (not shown) also reveals the chaotic phase 
(as can be anticipated from Fig.~\ref{fig:DynamicsH}).

Hence, the emergence of the dynamical ergodic regime for the staggered states $\ket{\psi_s}$ is indeed controlled by $\gamma$, in accordance with the energy scaling analysis of Sec.~\ref{sec:control}.
\subsection{Localized initial density}
\begin{figure*} 
	\centering
	\includegraphics[width=.98\textwidth]{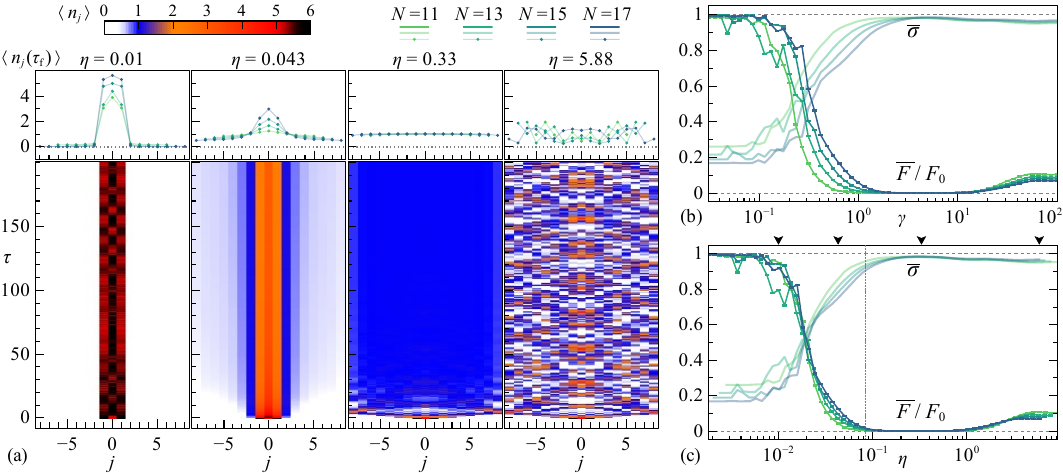}
	\caption{Dynamical observables for initially localized Fock states $\ket{\psi_\ell}$ with $\ell=3$ central sites populated as $\{\ket{434},\ket{454},\ket{555},\ket{656}\}$, respectively for  $N=L\in\{11,13,15,17\}$. Density plots in (a) show the evolution of on-site density expectation values, $\langle n_j(\tau)\rangle$, with tunneling time $\tau$ for $N=17$ and the four values of $\eta$ indicated [also marked by arrowheads in panel (c)], and are topped by the spatial density profiles at the final evolution time $\tau_\textrm{f}=200$. 
	 The time-averaged value of the density profile width [Eq.~\eqref{eq:sigmaW}], $\overline{\sigma}$, and of the density mean square deviation, $\overline{F}$ [Eq.~\eqref{eq:F}, scaled by $F_0\equiv F(\tau=0)$], in the interval $\tau\in[100,200]$ are shown for varying system size (color coded) as functions of $\gamma$ in panel (b) and of $\eta$ in panel (c). The vertical dotted line in (c) marks the threshold value $\eta^c_s=1/12$ from Eq.~\eqref{eq:etac} for $N\to\infty$.}
	\label{fig:DynamicsLoc}
\end{figure*}

Lastly, we consider the common experimental situation in which the bosonic density is initially contained in a restricted region of the system whence it expands in time \cite{Venzl2009}. 
We mimic this case by taking initial Fock states $\ket{\psi_\ell}$ at unit density with only $\ell=3$ central sites populated. 

The time evolution of the density distribution for $N=17$ and four values of the scaled tunneling strength $\eta$ [Eq.~\eqref{eq:eta}] can be observed in Fig.~\ref{fig:DynamicsLoc}(a). Like for the 
staggered configurations, strong
interaction ($\eta=0.01$) entails non-ergodic behaviour and persistence of the initial conditions in time. As the tunneling strength grows ($\eta=0.043$), the cloud starts registering an 
expansion in time, which has been drastically accelerated at $\eta=0.33$, for which full steady-state homogenization emerges after $\approx 30$ tunneling times. Upon approaching the 
weakly interacting limit ($\eta=5.88$), the 
evolution of the density profile displays a characteristic and recurrent interference pattern in time. 

These different behaviours are captured by the time-averaged values of the width $\sigma(\tau)$ and of the spatial density fluctuations $F(\tau)$, shown 
as functions of $\gamma$ in panel (b), and of $\eta$ in panel (c). Importantly, the distinct $\gamma$-regime of vanishing $\overline{F}$ and maximal width that one may 
associate with the chaotic phase witnesses a systematic shift toward larger values for increasing system size. In terms of $\eta$, however, the shift is offset, and the 
behaviour of $\overline{F}$ falls onto a common trend for all $N$ (up to, arguably, minor finite size effects). From Fig.~\ref{fig:DynamicsLoc}(c), the ensuing dynamical 
ergodic regime $0.08\lesssim \eta \lesssim 1$ positively correlates with the chaotic phase identified in 
our spectral analysis [cf.~Fig.~\ref{fig:JU_JUN_Comparison}(d)], 
confirming the validity of the analytical estimation of the threshold value from Eq.~\eqref{eq:etac}, $\eta^c_\ell=1/12$
$(N\to\infty)$. 

Therefore, the dynamics of initially localized bosonic clouds of fixed size and different particle number, or similar particle number but different size, for the same bare interaction strength, 
are not comparable. The potential emergence or absence of ergodicity in these cases is controlled by the rescaled tunneling strength $\eta$, as concluded from the spectral analysis in the 
previous section, and here confirmed.

\section{Conclusions}
\label{sec:conclusions}

We have established, in the paradigmatic Bose-Hubbard model, the perfect correspondence between the chaotic region identified from the spectral structure and the emergence of  ergodic dynamics,
if only the experimentalist is aware of the trajectory navigated along, across the parametrically evolving energy level structure, as seeded by the specific choice of the 
many-body initial state of a typical quench experiment.
The scaling of the energy expectation value with the number of bosons at fixed particle density permits to conclude whether a Fock initial state witnesses 
a chaotic phase controlled by the bare tunneling strength $\gamma=J/U$, or whether the persistence of chaos in the thermodynamic limit is determined by the 
parameter $\eta=J/UN$, scaled by the number of particles $N$. 
For the assessment of dynamical ergodic behaviour in the context of experimental studies with 
expanding atomic clouds, 
this, in particular,  implies that runs for 
different initial cloud sizes with similar $N$, for the same interaction strength (and hence identical values of $\gamma$), are not comparable:
As the onset of chaos is governed by $\eta$ and not $\gamma$ for these initial states, the $\gamma$-thresholds of chaotic behaviour are different for each cloud.
Furthermore, we have seen that the onset of chaos for any Fock initial state can be analytically 
estimated from the excess energy density above the ground state, together with
the 
crossover behaviour thereof from 
the strongly interacting to the non-interacting limit.

\begin{acknowledgments}
A.R. thanks Marcos Rigol for helpful discussions. 
The authors acknowledge support by the state of Baden-W\"urttemberg through bwHPC and the German Research Foundation (DFG) through Grants No.~INST 40/467-1 FUGG (JUSTUS cluster), No.~INST 40/575-1 FUGG (JUSTUS 2 cluster), and No.~402552777, and by Ministerio de Ciencia e Innovaci\'on/Agencia Estatal de Investigaci\'on (Spain) 
through Grant No.~PID2020-114830GB-I00. 
E.G.C.~acknowledges support from the Georg H.~Endress foundation.
This project (EOS 40007526) has received funding from the FWO and F.R.S-FNRS under the Excellence of Science (EOS) programme.
This research has made use of the high performance computing resources of the Castilla y Le\'on Supercomputing Center (SCAYLE, www.scayle.es), financed by the
European Regional Development Fund (ERDF), and of the CSUC (Consorci de Serveis Universitaris de Catalunya) supercomputing resources. 
We thankfully acknowledge RES resources provided by the Galician Supercomputing Center (CESGA) in FinisTerrae III to activity FI-2024-2-0027.
The supercomputer FinisTerrae III and its permanent data storage system have been funded by the Spanish Ministry of Science and Innovation, the Galician Government and the European Regional Development Fund (ERDF).
\end{acknowledgments}

\appendix 
\section{Energy expectation value for maximally mixed state}
\label{ap:mms}
Since the diagonal of Hamiltonian \eqref{eq:BHH} is solely determined by the interaction term, the energy expectation value for the maximally mixed state $\rho_\text{MM}=\mathbbm{1}/\HSD(N,L)$, where we recall that $\HSD(N,L)=\begin{psmallmatrix} N+L -1 \\ N \end{psmallmatrix}$, can be evaluated straightforwardly, 
\begin{align}
 \Tr(H\mms)&=\frac{1}{\HSD(N,L)}\frac{U}{2} \Tr \left[\sum_{j=1}^L\hat{n}_j (\hat{n}_j-1)\right] \notag\\
 &=\frac{U}{2} \left( \frac{1}{\HSD(N,L)}\sum_{j=1}^L\Tr \hat{n}_j^2  -N \right) \notag \\
 &=\frac{U}{2} \left( \frac{L}{\HSD(N,L)} \sum_{\|\vec{n}\|_1=N} n_1^2  -N \right) \notag \\
 &=\frac{U}{2} \left( L \sum_{k=1}^N k^2 \frac{\HSD(N-k,L-1)}{\HSD(N,L)}  -N \right) \notag \\
 &=U\frac{N(N-1)}{L+1}.
\end{align}
For the case of constant bosonic density, $n=N/L$,
\begin{equation}
 \Tr(H\mms)=U\left[ nN -n(n+1) +O(N^{-1})\right],
\end{equation}
i.e., the energy expectation value scales linearly with particle number, $\Tr(H\mms)\simeq UnN$, as the thermodynamic limit is approached. 

\section{Chebyshev expansion technique for time evolution}
\label{ap:chebyshev}
The time evolution of a generic initial state $\ket{\psi_0}$, induced by the time-independent Bose-Hubbard Hamiltonian, can be numerically implemented by expanding the corresponding time evolution operator $\mathcal{U}(\tau)=e^{-i(H/J)\tau}$, with $\tau\equiv Jt/\hslash$, in terms of Chebyshev polynomials of the first kind $T_n(x)$, $n\geqslant 0$ \cite{Weisse2008}. A rescaled Hamiltonian $\tilde{H}$ must be defined from $H/J=a\tilde{H}+b\mathbbm{1}$, with 
\begin{equation}
 a=(E_\text{max}-E_\text{min})/2J, \quad  b=(E_\text{max}+E_\text{min})/2J,
\end{equation}
to ensure that $\spec(\tilde{H})\in[-1,1]$. The forward evolution from time $\tau$ to $\tau+\Delta\tau$ can then be written as 
\begin{equation}
 \mathcal{U}(\Delta\tau)\ket{\psi(\tau)}\simeq e^{-ib\Delta\tau}\left[c_0\ket{v_0(\tau)} + 2\sum_{n=1}^M c_n \ket{v_n(\tau)} \right],
 \label{eq:Uexp}
\end{equation}
where the coefficients are determined by Bessel functions, $J_n(x)$, 
\begin{equation}
 c_n\equiv (-i)^n J_n(a\Delta\tau),
\end{equation} 
and $M$ denotes the chosen cut-off for the expansion. The vector states in the series follow from the recursion relation of the Chebyshev polynomials, 
\begin{subequations}
\begin{align}
 \ket{v_0(\tau)} &\equiv \ket{\psi(\tau)}, \\
 \ket{v_1(\tau)} & \equiv T_1(\tilde{H})\ket{v_0(\tau)} =\tilde{H} \ket{\psi(\tau)}, \\
 \ket{v_{n+1}(\tau)} &\equiv T_{n+1}(\tilde{H})\ket{v_0(\tau)} = 2\tilde{H}\ket{v_n(\tau)}- \ket{v_{n-1}(\tau)}.
\end{align}
\label{eq:recurV}
\end{subequations}
Given a value of $a\Delta\tau$, the coefficients $c_n$ decay faster than exponentially with $n$ for large $n$, and hence an accurate expansion may be achieved with a reasonable number of terms, although, for a targeted precision, the number of terms will increase for larger spectral width $a$ and/or chosen time step $\Delta\tau$. 
The implementation of $\eqref{eq:Uexp}$, as can be seen in Eqs.~\eqref{eq:recurV}, only requires matrix-vector multiplications, which can be efficiently parallelized \cite{petsc-user-ref,petsc-efficient,petsc-web-page,slepc}. 

We carry out dynamics without imposing any truncation of the maximum occupation number in the Fock basis, and always take initial Fock states that exhibit parity symmetry to reduce the dynamically available Hilbert space. We consider systems at unit density up to $L=17$ with Hilbert subspace $\mathcal{H}^+$ of size $\mathcal{D}^+\simeq 5.8\times 10^8$. 
The expansion cut-off $M$ is chosen to keep all terms with coefficients obeying $|c_n|\geqslant 10^{-12}$ for any $\gamma=J/U$. In all cases, we have checked that the chosen precision is enough to ensure convergence of the calculated observables.

%
\end{document}